\documentstyle[preprint,floats,tighten,aps,graphicx]{revtex}

\newif\ifpdf
\ifx\pdfoutput\undefined
\pdffalse 
\else
\pdfoutput=1 
\pdftrue
\fi

\def\SppP{{\cal {P\!\!\!\!\hspace{0.04cm}\slash}}_\perp}

\def\nslash{n\!\!\!\slash}
\def\bnslash{\bar n\!\!\!\slash}

\def\Aslash{A\!\!\!\slash}
\def\delP#1{}
\def\OMIT#1{}

\newcommand{\bilin}[3]{{\bar #1\, #2 \, #3}}
\newcommand{\nn}{\nonumber} 
\newcommand{\bn}{\bar n}

\newcommand{\bnP}{\bar {\cal P}}

\newcommand{\W}{W}

\newcommand{\mcdot}{\!\cdot\!}
\newcommand{\Ub}{{\cal U}}

\preprint{\vbox{ \hbox{UCSD/PTH 01-09}   \hbox{hep-ph/0107001} }}
\title{Invariant Operators in Collinear Effective Theory}
\author{Christian W. Bauer\footnote{bauer@einstein.ucsd.edu} and 
Iain W. Stewart\footnote{iain@schwinger.ucsd.edu} \\[20pt]}
\address{ \vbox{\vskip 0.truecm}  \tighten 
Physics Department, University of California at San Diego, La Jolla, CA 92093 
}
\begin{document}

\ifpdf
\DeclareGraphicsExtensions{.pdf, .jpg}
\else
\DeclareGraphicsExtensions{.eps, .jpg, .ps}
\fi

\maketitle
\begin{abstract}

We consider processes which produce final state hadrons whose energy is much
greater than their mass.  In this limit interactions involving collinear
fermions and gluons are constrained by a symmetry, and we give a general set of
rules for constructing leading and subleading invariant operators.  Wilson
coefficients $C(\mu,\bnP)$ are functions of a label operator $\bnP$, and do not
commute with collinear fields. The symmetry is used to reproduce a two-loop
result for factorization in $B\to D\pi$ in a simple way.

\end{abstract}
\tighten
\newpage

\section{Introduction}

For processes involving the production of a highly energetic hadron, $E_H\gg
m_H$, a large number of quarks and gluons typically move close to a light cone
direction $n^\mu$.  The QCD dynamics of these collinear particles is easiest to
describe with light cone coordinates $p = (p^+, p^-, p_\perp)$, where $p^+\!=\!
n\cdot p\,$, $p^-\!=\!\bn\cdot p\,$.\footnote{For motion in the $+z$ direction,
$n^\mu\!=\!(1,0,0,1)$ and $\bn^\mu\!=\!(1,0,0,-1)$.} The size of the momentum
components of a collinear particle are very different, $p^- \sim Q$ is large,
while $p^\perp\sim Q\lambda$ and $p^+\sim Q\lambda^2$ are smaller by an amount
dictated by a parameter $\lambda\ll 1$. In problems with well separated scales
it is often profitable to construct an effective theory where the dynamics
associated with the large momentum scale are integrated out.  In
Refs.~\cite{bfl,bfps} a collinear-soft effective theory with a power counting in
$\lambda$ was constructed to describe the production of fast moving hadrons in
heavy-to-light decays such as $B\to X_s\gamma$ and $B\to \rho\ell\nu$.  In this
paper we clarify and generalize the construction of effective theories with
collinear particles.

\begin{figure}[h!]
  \centerline{ \includegraphics[width=4in]{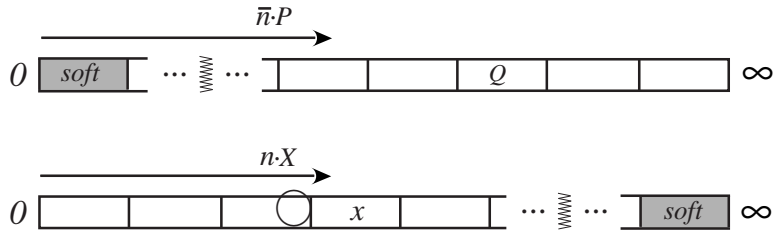} } 
\vskip 0.4cm
 \caption{Typical momenta/positions for soft particles (shaded bins), and
  collinear particles (white  bins). \label{fig_grids}}
\end{figure}

We wish to describe collinear quarks and gluons produced by an interaction
specifying a non-collinear reference frame. The latter frame can be set by a
current, heavy meson, or set of soft particles, and makes it impossible to
eliminate all large collinear momenta by a boost. We want to describe this
system by a local field theory. The problem is that interactions between nearly
{\em onshell} soft and collinear particles are non-local in the $x^+$
direction. For instance, a collinear gluon can not interact with a soft fermion
without taking it far off its mass shell.  In the limit of very large $Q$, the
situation is pictured in Fig.~\ref{fig_grids}. Discretizing the large momentum
of collinear particles, they live in bins far from $\bn\mcdot P=0$, while soft
particles populate the $\bn\mcdot P=0$ bin.  In position space the collinear
particles are short distance and are in bins near $n\cdot X=0$, while the soft
particles are effectively in a bin infinitely far away.  To construct the
effective field theory all offshell fluctuations are integrated out. This gives
operators which are non-local in the position $n\mcdot X$ or momentum $\bn\mcdot
P$, but are local in the other coordinates.

The situation becomes interesting when we consider what form gauge invariance
takes in this effective theory.  Consider a bilinear quark operator $\bar\psi_c
\psi_s$, where $\psi_s$ is soft and $\psi_c$ is collinear. If we want the
collinear particle at $X\!=\!x$ and the soft particle at $X\!=\!\infty$ then
gauge invariance requires them to be connected by a Wilson line $W$, so the
operator is $\bar\psi_c(x) W(x,\infty) \psi_s(\infty)$.\footnote{Similar eikonal
lines appear in the language of jet physics~\cite{collins}.}  In QCD a gauge
transformation takes $W(x,y)\to e^{i\alpha(x)} W(x,y) e^{-i\alpha(y)}$. Now,
consider a ``collinear'' gauge transformation which has no support at $\infty$,
so that $e^{-i\alpha(\infty)}\!=\!1$. The change of collinear quarks under this
transformation is what requires the presence of collinear gluons. A collinear
gluon cannot interact with soft particles without taking them far offshell, so
$\psi_s(\infty)$ does not transform. The transformation simply connects
$\bar\psi_c(x)$ to $W(x,\infty)$. In momentum space, the collinear gluon fields
in $W$ and the collinear quark field $\psi_c$ are labelled by their large
momenta, corresponding to the bins in Fig.~\ref{fig_grids}. From matching, the
bilinear quark operator can be multiplied by a complicated short distance
function of these momenta $C(\bn\cdot p_i)$.  However, the collinear gauge
symmetry greatly restricts this function, and gives the effective theory
framework predictive power.

In section~\ref{sect_ops} we set up a formalism which makes it simple to
construct invariant field operators.  This is done with the help of an operator
$\bnP$ which acts in the label space of collinear fields. This insight allows a
simple set of rules to be derived, from which the most general allowed operators
involving collinear fields can be constructed. This includes both leading as
well as power suppressed contributions. For simplicity we focus on collinear
gluons, although including soft gluons does not alter any of the results
presented.

In section~\ref{sect_eg} we give two simple examples of how the formalism is
useful. We construct a compact gauge invariant expression for the collinear
quark kinetic term, and discuss the lowest order heavy-to-light weak currents in
the effective theory.

In section~\ref{sect_bdpi} we discuss a non-trivial example by constructing
effective theory operators for the decays $B^0\to D^+\pi^0$ and $B^-\to
D^0\pi^-$. The authors of Ref.~\cite{polwise,bbns} proposed that at leading
order in $1/E_\pi$ and $1/m_{b,c}$, the terms that break naive factorization for
these decays are computable perturbatively.  They put forth a generalized notion
of factorization, in which the matrix element of a four-quark operator can be
written as the product of a $B \to D$ form factor, and the convolution of a
short distance coefficient with the light cone wave function of the pion. In
Ref.~\cite{bbns2} it was shown that at two-loops the only non-zero long distance
contributions which could spoil generalized factorization come from the
hard-soft and hard-collinear regions of phase space. However, it was
shown~\cite{bbns2} that the former give a one-loop hard coefficient times the
one-loop correction to the $B \to D$ form factor, while the latter reduce to a
convolution of the one-loop short distance coefficient and the one-loop pion
wavefunction~\cite{BL}.  Here we show that the allowed operators in the
effective theory reproduce the hard-collinear contribution in a simple way, and
also extend the resulting convolution to include the short distance coefficient
at an arbitrary order in perturbation theory.

\section{Invariant Operators} \label{sect_ops}

To explore the nature of collinear interactions we begin by taking a free field
$\phi(x)$ with the mode expansion written with a relativistic measure
\begin{eqnarray} \label{phi1}
  \phi(x) &=& \int \frac{d^4P}{(2\pi)^3} \:\delta(P\,^2)\, \theta(P^0)\, \bigg[ 
    {\cal X}(P)\: a\mbox{\footnotesize $(P)$}\, e^{-iP\cdot x} + 
    {\cal Y}(P)\: b^\dagger\mbox{\footnotesize $(P)$}\, e^{iP\cdot x} \bigg] 
  \equiv \phi^+(x) + \phi^-(x)    \,. 
\end{eqnarray}
Here $a\mbox{\footnotesize $(P)$}$ annihilates particles,
$b^\dagger\mbox{\footnotesize $(P)$}$ creates antiparticles, and ${\cal X}(P)$
and ${\cal Y}(P)$ take values depending on the representation of the Lorentz
group to which $\phi$ belongs (spinors, polarization vectors, etc.).  When $P$
is collinear it scales as $P= (P^+, P^-, P^\perp)\sim Q(\lambda^2,1,\lambda)$
where $Q$ is a large scale and $\lambda$ is small. To define the effective
theory the order $\lambda^0$ momenta must be singled out, since it is these
momenta we wish to expand about. Therefore we let $P=p+k$ where $p$ includes the
$\lambda^0$ part of $P^-$ and the $\lambda$ part of $P^\perp$, leaving $k^\mu
\sim \lambda^2$.  We include the $P^\perp$ momenta in $p$ mostly as a matter of
convenience; making these momenta explicit simplifies the power
counting~\cite{bfps}.  The large phases in Eq.(\ref{phi1}) can now be removed by
defining effective theory fields $\phi^\pm_{n,p}$ with labels $n$ and $p$ (where
$n$ defines the light cone vector the particle is moving along). Thus,
\begin{eqnarray} \label{phi2}
  \phi^+(x) &\to & \sum_p e^{-ip\cdot x} \phi^+_{n,p}(x) \,,\qquad\quad
  \phi^-(x) \to \sum_p e^{ip\cdot x} \phi^-_{n,p}(x) \,, \\
  \phi^+_{n,p}(x) &=&  \int \frac{d^4k}{(2\pi)^3}\,  \theta(\bn\mcdot p)\:
    \delta( n\mcdot k\,\bn\mcdot p+p_\perp^2)
    \: \tilde {\cal X}_p(k) \: a_p(k)\: e^{-ik\cdot x} \,, \nn \\
  \phi^-_{n,p}(x) &=&  \int \frac{d^4k}{(2\pi)^3}\,  \theta(\bn\mcdot p)\:
    \delta( n\mcdot k\,\bn\mcdot p+p_\perp^2)
    \: \tilde {\cal Y}_{p}(k) \: b^\dagger_{p}(k)\: e^{ik\cdot x}\,.   \nn
\end{eqnarray} 
Since $n\cdot p=0$, only particles with large momentum $p$ in the $n^\mu$
direction are described. The field $\phi^+_{n,p}$ destroys such particles and
$\phi^-_{n,p}$ creates the corresponding anti-particles. All fields are created
or destroyed with residual momentum $k$. To simplify the notation we define
\begin{eqnarray}
  \phi_{n,p}(x) \equiv \phi^+_{n,p} + \phi^-_{n,-p} \,, \label{newphi}
\end{eqnarray}
so that $\phi_{n,p}$ with positive (negative) label $\bn\mcdot p$ denotes the
particle (antiparticle) field.\footnote{In Ref.~\cite{bfps} this sign was used
for quarks, but collinear gluon fields were labelled by $-p$.} The ${\cal
O}(\lambda^0)$ momentum labels on fields correspond to the bins in
Fig.~\ref{fig_grids}.

In Eq.~(\ref{phi2}) we have expanded about the large light cone momentum which
is the meaning of the tilde in the effective theory functions $\tilde {\cal
X}_p$ and $\tilde {\cal Y}_p$. For collinear fermions, the field $\xi_{n,p}$
satisfies $\nslash \xi_{n,p}=0$, and therefore has only two components.  For
collinear gluons the effective theory fields satisfy $(A^\mu_{n,q})^* =
A^\mu_{n,-q}$.

If the labels $p$ were conserved quantum numbers in this large energy limit of
QCD, the remainder of the discussion would be simple. The situation would be
much like the velocity label $v$ on the effective theory fields in heavy quark
effective theory (HQET)~\cite{Georgi}. However, in our case the interaction of
infrared degrees of freedom still cause order one changes in the labels due to
the peculiar nature of light cone coordinates; since $P\,^2 \simeq p^- k^+$,
even a small change in $P\,^2$ can change the large $\lambda^0$ momentum $p^-$
by an amount of order one.  For this reason it might seem that the decomposition
in Eq.~(\ref{phi2}) is not very useful. For instance, in loop integrals we will
still need to perform the sum over the large $p^-$ momenta.

The key observation is that after expanding, the remnant of gauge symmetry
provides important restrictions on the allowed form of label changing
operators. It is these restrictions that make the effective theory predictive,
and determining the corresponding rules is the goal of this paper.

To proceed, we make a brief diversion to define an operator $\bnP$ which acts
on products of effective theory fields. When acting on collinear fields,
$\bnP$ gives the sum of large labels on fields minus the sum of large labels on
conjugate fields. Thus, for any function $f$
\begin{eqnarray}
 && f(\bnP) \Big(\phi^\dagger_{q_1} \cdots \phi^\dagger_{q_m} 
 \phi_{p_1} \cdots \phi_{p_n}\Big) \\ 
 && \qquad = f(\bn\mcdot p_1\!+\!\ldots\!+\!\bn\mcdot p_n\! 
 -\!\bn\mcdot q_1 \!-\!\ldots\!-\!\bn\mcdot q_m)  
 \Big(\phi^\dagger_{q_1} \cdots \phi^\dagger_{q_m} \phi_{p_1} \cdots 
 \phi_{p_n}\Big) \,. \nn
\end{eqnarray}
The operator $\bnP$ has mass dimension $1$, but power counting dimension
$\lambda^0$. The conjugate operator $\bnP^\dagger$ acts only to its left and
gives the sum of labels on conjugate fields minus the sum of labels on fields.
For convenience we also define an operator ${\cal P}_\perp^\mu \sim \lambda$
which produces the analogous sum of order $\lambda$ labels, and we let ${\cal
P}^\mu = \frac12 n^\mu \bnP + {\cal P}_\perp^\mu$.

The power counting for collinear quark fields is $\xi_{n,p}\sim \lambda$, while
for collinear gluon fields $(A^+_{n,q}, A^-_{n,q}, A^\perp_{n,q})\sim
(\lambda^2,1,\lambda)$~\cite{bfps}.  Since $\bn\mcdot A_{n,q}\sim \lambda^0$ an
operator at a given order can contain any number of these gluons. The utility of
${\cal P}^\mu$ is that even in an operator with an arbitrary number of fields
the large phases can all be pulled to the front of the operator. This is because
these phases can be commuted through derivatives,
\begin{eqnarray}
 i\partial^\mu\: e^{-ip\cdot x} \phi_{n,p}(x) = e^{-ip\cdot x}({\cal
  P^\mu}+i\partial^\mu) \phi_{n,p}(x) \,,
\end{eqnarray}
and the overall phase then simply enforces label conservation.  The remaining
derivatives on fields are order $\lambda^2$.  Furthermore, using these operators
we no longer need to write explicit sums over labels.  Since $\bnP^\mu$ commutes
with the sums, we can adopt a convention where all field labels are summed over,
$\sum_p \phi_{n,p} \to \phi_{n,p}$. For e.g., $\sum_p \bnP \phi_{n,p}= \bnP
\sum_p\phi_{n,p} \to \bnP \phi_{n,p}$. From now on labels are always summed
over, unless otherwise stated. This leads to our first two rules which are in
some manner notational:
\begin{enumerate}

 \item[1)] Changes of variable on the labels of fields are allowed. (The original
   summation variable was after all arbitrary.)\\[-12pt]

 \item[2)] In writing Feynman rules the label momenta must be conserved. 

\end{enumerate} 
An example of using 1) is: $\bn\mcdot t\: \phi_{n,r}\,\phi_{n,s-r}\,\phi_{n,t}
=\bn\mcdot u\: \phi_{n,r}\, \phi_{n,s-r}\, \phi_{n,u} = \bn\mcdot u\:
\phi_{n,r}\, \phi_{n,s}\, \phi_{n,u}$.  Taking matrix elements and using rule 2)
each term gives the same Feynman rule.  

Now consider a non-abelian gauge transformation $U(x)=\exp[{i\alpha^A(x)T^A}]$
with all of its support over a set of collinear momenta. Much like in
Eq.~(\ref{phi2}) it is useful to decompose this collinear transformation into a
sum over these collinear momenta
\begin{eqnarray} \label{U}
   U(x) = \sum_Q e^{-iQ\cdot x} \Ub_Q \,.
\end{eqnarray}
Expanding the QCD gauge transformation and using rules 1) and 2) yields
particularly simple transformation rules for collinear fermions and gluons (for
fixed labels $p$ and $q$)
\begin{eqnarray} \label{gt}
  \xi_{n,p} &\to& \Ub_{p-Q}\ \xi_{n,Q} \,, \qquad
  A^\mu_{n,q} \to \Ub_Q\: A^\mu_{n,R}\: \Ub^\dagger_{Q+R-q} 
  + \frac{1}{g}\, \Ub_{Q} \Big[ {\cal P}^\mu\: \Ub^\dagger_{Q-q} \Big] \,.
\end{eqnarray}
The square bracket indicates that ${\cal P}^\mu$ acts only on terms within the
bracket.  In general ${\cal P}^\mu {\cal O}=[{\cal P}^\mu {\cal O}]+{\cal
O}{\cal P}^\mu$.  For an Abelian gauge group Eq.~(\ref{gt}) agrees with
Ref.~\cite{bfps}.

It is convenient to define a function $\W$ of $\bn\mcdot A_{n,q}$ such that
$W^\dagger\: \xi_{n,p}$ is invariant under the transformation in
Eq.~(\ref{gt}). Consider
\begin{eqnarray} \label{W}
  \W &=& \bigg[ 
  \lower7pt \hbox{ $\stackrel{\sum}{\mbox{\scriptsize perms}}$ }
  \!\! \exp\bigg(-\!g\,\frac{1}{\bnP}\ \bn\mcdot A_{n,q} \bigg) \bigg] \,, 
  \qquad\quad \W^\dagger = \bigg[ 
  \lower7pt \hbox{ $\stackrel{\sum}{\mbox{\scriptsize perms}}$ }
  \!\!  \exp\bigg(-\!g\,\bn\mcdot A^*_{n,q}\
  \frac{1}{\bnP^\dagger}\ \bigg) \bigg] \,,
\end{eqnarray}
which satisfy $W^\dagger W=1$. In the expansion of the exponential the $1/\bnP$
acts to the right on all gluon fields in the square brackets, and we sum over
permutations of the gluon fields. In Feynman rules the $1/n!$ in this expansion
cancels the $n!$ from the choices for contracting the gluons. It is easy to show
that $\W$ satisfies the linear equation
\begin{eqnarray} \label{Weom}
   \bigg[ \Big(  \bnP \delP{q} +  g\bn\mcdot A_{n,q} \Big) \W \bigg] = 0 \,.
\end{eqnarray}
With a given boundary condition the solution of this equation is
unique. Now consider
\begin{eqnarray}
  &&  \bigg[ \bigg(\bnP\:\delP{q} + g\bigg\{ \Ub_Q\: \bn\mcdot A_{n,R}\: 
  \Ub^\dagger_{Q+R-q} + \frac{1}{g} \Ub_{Q} 
    \Big[ {\bnP}\, \Ub^\dagger_{Q-q} \Big] \bigg\} \bigg) \Ub_T \W \bigg]  \nn\\
 && = \bigg[ \bigg( \Ub_{T} \, \{ \bn\mcdot T + \bnP \} 
 + g \Ub_Q\, \bn\mcdot A_{n,R}\, \Ub^\dagger_{-q}\,\Ub_T
 + \bn\mcdot (q-Q)\: \Ub_{Q}\,\Ub^\dagger_{Q-q}\, \Ub_T \bigg) \W \bigg] \nn  \\
 && = \bigg[ \bigg( \Ub_{T} \, \{ \bn\mcdot T +\bnP \}
 + g \Ub_Q\, \bn\mcdot A_{n,R} 
 - \bn\mcdot Q\: \Ub_{Q}\, \Ub^\dagger_{-q}\, \Ub_T \bigg) \W  \bigg] \\
 && = \bigg[ \Ub_T (\bnP + g\bn\mcdot A_{n,R} ) \W \bigg] = 0\,. \nn
\end{eqnarray}
In the second line we have used (for fixed $p$) $\bnP \phi_{n,p} = [\bnP
\phi_{n,p}] +\phi_{n,p}\bnP = \phi_{n,p} (\bn\mcdot p+\bnP)$. To obtain the
third and fourth lines unitarity of the gauge transformation $\Ub^\dagger_{P+r}\
\Ub^{\phantom{*}}_{P+r'} = \delta_{r,r'}$ (with fixed $r$, $r'$) was used.  We
see that $\Ub_T \W$ is a solution of the linear equation with $(\bnP +
g\bn\,\mcdot A_{n,q})$ transformed using Eq.~(\ref{gt}).  Thus, from uniqueness
the transformation is
\begin{eqnarray}\label{Wtrafo}
  \W \to \Ub_T \W \,,
\end{eqnarray}
which makes $W^\dagger \xi_{n,p}$ invariant under a collinear gauge
transformation. Actually, $\bar\xi_{n,p} W \delta_{\bnP^\dagger,P}$ is the
Fourier transform of the quantity $\bar\xi_n(x) W(x,\infty)$ discussed in
section I ($\delta$ is a Kronecker of delta). Eq.~(\ref{Weom}) is the parallel
transport equation for the momentum space Wilson line $W$. Invariant operators
with other fields can also be constructed, for instance the order $\lambda$
quantity $( \SppP + g \Aslash_{n,q}^\perp )$ acts like a covariant derivative.
We have (for fixed $p$)
\begin{eqnarray}
  \SppP \, \xi_{n,p}+ g \Aslash_{n,Q}^\perp \, \xi_{n,p-Q} 
   \ \to\ \Ub_{p-T} ( \SppP \, \xi_{n,T}+ g \Aslash_{n,Q}^\perp \, \xi_{n,T-Q}) 
  \,.
\end{eqnarray}

Wilson coefficients in an effective theory depend on the large scales, and
therefore in our case on the $\bn\mcdot p$ labels on the fields. Since the
collinear theory does not conserve labels, the Wilson coefficients can depend on
$\bnP$ and $\bnP^\dagger$ and have to be inserted {\em between} the fields in an
operator. In fact, invariance under Eq.~(\ref{gt}) ensures that {\em only} the
linear combination of labels picked out by $\bnP$ and $\bnP^\dagger$ can appear
in these coefficients.  For a function $f$, Eq.~(\ref{Weom}) can be used to prove
that
\begin{eqnarray} \label{ftof}
  f(\bnP +g\bn\mcdot A_{n,q}) =  \W\ f(\bnP)\ \W^\dagger \,,
\end{eqnarray}
which guarantees that gauge invariant combinations of $\bn\mcdot A_{n,q}$ fields
only appear in $W$.  We conclude that the most general allowed operators must
satisfy two more rules:\\[-15pt]
\begin{enumerate}
 \item[3)] Only operators invariant under the transformations in
 Eq.~(\ref{gt}) are allowed.\\[-15pt]

 \item[4)] Wilson coefficients are functions of the label operators $\bnP$ and
 $\bnP^\dagger$ and must be inserted in all possible locations in a field
 operator that are consistent with 3).
\end{enumerate}


\section{Simple examples} \label{sect_eg}

Our first example of the constraints imposed by the collinear gauge
transformation is the kinetic term for collinear quarks.  In Ref.~\cite{bfps} an
action was constructed which gave the free propagator $i(\nslash/2)/[n\cdot
k+p_\perp^2/(\bn\cdot p)]$ for collinear quarks, as well as their leading
interactions with collinear gluons.  However, because of the infinite number of
couplings to $\bn\mcdot A_{n,q}$ gluons a closed form expression for the ${\cal
O}(\lambda^0)$ Lagrangian was not found. Such a closed gauge invariant form is
easily deduced in terms of the label operators\footnote{The terms in
Eq.~(\ref{Lc}) are protected from obtaining an anomalous dimension by
normalizing the free quark kinetic term.  If soft gluons are included then in
Eq.~(\ref{Lc}) the term $n\cdot i\partial\to n\cdot iD$ making the first term in
the action identical to the LEET Lagrangian~\cite{leet}.}
\begin{eqnarray} \label{Lc}
  {\cal L} &=&  \Bigg[ \bar\xi_{n,p'}\:  \bigg\{
  n\!\cdot i{\partial} + g n\mcdot A_{n,q} \, 
  + \Big( \SppP  + g \Aslash_{n,q}^\perp\Big)\, \W\ \frac{1}{\bnP}\ \W^\dagger\,
   \Big( \SppP  + g \Aslash_{n,q'}^\perp\Big) \bigg\}
  \frac{\bnslash}{2}\, \xi_{n,p} \Bigg] \,,
\end{eqnarray}
where the fields are all functions of $x$.  To make the dimensions consistent
two powers of $\Big( \SppP + g \Aslash_{n,q}^\perp\Big)$ must be accompanied by
a $1/\bnP$. The $1/\bnP$ sits between the fields; it can not appear in front
since the sum of $\bn\cdot p$ momenta is zero by label conservation. Now, the
gauge transformation on collinear fermions induces factors that do not commute
with functions of $\bnP$, requiring the presence of the $\W^\dagger$ and $\W$ to
give an invariant operator. Thus, the factors to the left and right of the
$1/\bnP$ are separately invariant.  The order $\lambda^0$ interactions with
collinear quarks include all components of $A_{n,q}^\mu$ and reproduce those in
Ref.~\cite{bfps}. Expanding the $\W's$ in $g$ gives an infinite set of couplings
to $\bn\cdot A_{n,q}$ gluons.

Next, consider the example of currents with one heavy and one collinear quark
field. These currents are relevant for $B$ decays to highly energetic light
hadrons such as $B\to K^*\ell^+\ell^-$ and $B\to\pi\ell\nu$. For example, they
give the soft form factor for heavy to light decays, in which only soft gluons
with momenta $\sim\!\!\Lambda_{\rm QCD}$ interact with the spectator in the $B$.
The exchange of a hard gluon with the spectator can induce operators with more
than one collinear quark field, but these will not be considered here.  The
heavy-to-light current at leading order in the effective theory is
\begin{eqnarray} \label{Jbpi}
  J^\Gamma = \bigg[ \bar\xi_{n,p}\: W\:\Gamma \: h_v \ 
  C^\Gamma\!(\mu,\bnP^\dagger) \bigg] \,,
\end{eqnarray}
where $h_v$ is the heavy quark HQET field, $\Gamma$ is one of the four
independent spin structures\cite{french,bfps}, and the Wilson coefficient
$C^\Gamma$ depends on $m_b$, $\mu$, and $\bnP^\dagger$.  Invariance under
collinear gauge transformations forces the combination $\bar\xi_{n,p} W$ to be
kept together, and $C^\Gamma$ {\em only} appears as an overall factor. Since
$\bnP^\dagger$ does not act on $h_v$ it picks out the sum of large momenta from
the collinear quark and gluons, so Eq.~(\ref{Jbpi}) reproduces the result in
Ref.~\cite{bfps}. Since $C^\Gamma$ is only a function of the total large
momentum it will never depend on a collinear loop momentum and always factors
outside of loops with collinear gluons.

The effective theory pion state $|\pi_{n,p_\pi}\rangle$ is labelled by its large
light cone momentum $p_\pi^\mu=E_\pi n^\mu$. For a function $f$, the operator
$\bnP$ acts on this state so that (for fixed $p_\pi$)
\begin{eqnarray}
  f(\bnP) \Big| \pi_{n,p_\pi} \Big\rangle 
  = f(\bn\cdot p_\pi) \Big| \pi_{n,p_\pi} \Big\rangle 
  = f(2 E_\pi) \Big| \pi_{n,p_\pi} \Big\rangle \,.
\end{eqnarray}
Therefore, in the effective theory the matrix element of the current 
\begin{eqnarray}
 && \Big\langle \pi_{n,p_\pi} \Big| \Big[ \bar\xi_{n,p}\: W\:\Gamma \: h_v\
  C^\Gamma\!(\mu,\bnP^\dagger) \Big] \Big| H_v \Big\rangle \nn\\
 &&\qquad = \bigg[ \Big\langle \pi_{n,p_\pi} \Big|C^\Gamma\!(\mu,\bnP^\dagger) 
 \bigg]
 \bar\xi_{n,p}\: W\:\Gamma \: h_v\  \Big| H_v \Big\rangle 
 -\Big\langle \pi_{n,p_\pi} \Big|  \bar\xi_{n,p}\: W\:\Gamma \: h_v\
 \bigg[ C^\Gamma\!(\mu,\bnP)  \Big| H_v \Big\rangle \bigg] \nn\\
 && \qquad =  C^\Gamma\!(\mu,2E_\pi)\: \Big\langle \pi_{n,p_\pi} 
 \Big| \bar\xi_{n,p}\: W\:\Gamma \: h_v \Big| H_v \Big\rangle \,,
\end{eqnarray}
where we used momentum conservation, $\langle \pi_{n,p_\pi} | J^\Gamma\, | H_v
\rangle\, \bnP^\dagger =0$, and the fact that the heavy meson state has no large
label.  This explains how the Wilson coefficient ends up depending on the scale
$E_\pi$, even though this is not the energy of any individual parton in the
current.


\section{The decays $\bar B^0\to D^+\pi^-$ and $B^-\to D^0\pi^-$} 
\label{sect_bdpi}

The kinematics of $B\to D\pi$ dictate that the energy of the pion is large
compared to its mass. The quarks in the pion can therefore be described by
collinear fermions in the effective theory.  In the full theory the relevant
weak Hamiltonian at $\mu_0=m_b$ is given in terms of four-quark operators ${\cal
O}_{\bf 0,8}(m_b)$
\begin{eqnarray}
  H_W &=& {4G_F\over \sqrt{2}} V_{cb}^{\phantom{*}}\, V_{ud}^* 
  \bigg[C^{\rm full}_{\bf 0} \left(m_b\right) O_{\bf 0} (m_b)\label{1.5.10} 
  + C^{\rm full}_{\bf 8} \left( m_b \right) O_{\bf 8} (m_b) \bigg],\\[3pt]
  O_{\bf 0}(m_b) &=& ( \bilin {c}{ \gamma^\mu P_L }{ b}) 
   (\bilin{d}{\gamma_\mu P_L }{ u} ), \qquad
  O_{\bf 8}(m_b) = ( \bilin {c}{\gamma^\mu P_L T^A}{ b}) 
   (\bilin {d}{ \gamma_\mu P_L T^A}{ u} )\,, \nn
\end{eqnarray}
where $P_L=(1-\gamma_5)/2$. We wish to match $O_{\bf 0,8}(m_b)$ onto operators
in the effective theory. At leading order, the most general operators which
satisfy the rules in section II are
\begin{eqnarray}\label{fourfermi_effective}
 Q_{\bf 0}^{\{1,5\}} &=& \Big( \bar{h}_{v'}^{(c)} \: 
  \frac{\nslash}{2}\{1,\gamma_5\}\: h^{(b)}_{v}\Big) 
  \Big(\bar{\xi}_{n,p'}^{(d)}\: W \, 
   C_{\bf 0}^{\{1,5\}}(\mu,v \!\cdot\!v',\bnP, \bnP^\dagger)
  \,\frac{\bnslash}{2}P_L \, W^\dagger \xi^{(u)}_{n,p} \Big) \,,\nn\\[5pt]
 Q_{\bf 8}^{\{1,5\}} &=& \Big( \bar{h}_{v'}^{(c)} \: 
  \frac{\nslash}{2}\{1,\gamma_5\}\, T^A\: h^{(b)}_{v}\Big) 
  \Big(\bar{\xi}_{n,p'}^{(d)}\: W\, 
  C_{\bf 8}^{\{1,5\}}(\mu,v \!\cdot\!  v',\bnP, \bnP^\dagger)\,
  \frac{\bnslash}{2}P_L\, T^A \, W^\dagger \xi^{(u)}_{n,p} \Big) \,.
\end{eqnarray}
The combinations $\bar{\xi}_{n,p'} W$ and $W^\dagger \xi_{n,p}$ are separately
invariant under Eq.~(\ref{gt}), so the most general operator involves a Wilson
coefficient inserted between these two terms.  (An overall $C(\bnP)$ can be
absorbed into a $C(\bnP,\bnP^\dagger)$ on the inside.) In general, $C_{\bf 0,8}$
depend on $\mu$, $m_{b}$, $m_{c}$, $v \mcdot v'$, and the operators $\bnP_{\pm}
= \bnP^\dagger \pm \bnP$.

In Ref.~\cite{bbns2} it was shown that at two-loops all long distance
contributions to $B\to D\pi$ can be absorbed into the $B\to D$ form factor and
the light-cone pion wavefunction. Graphs with only collinear and soft ($k^\mu\ll
Q$) gluons which could break factorization were shown to sum up to zero. Mixed
hard-soft graphs reduce to the one-loop hard coefficient times the one-loop
correction to the $B\to D$ form factor. Finally, mixed hard-collinear two-loop
graphs also contribute, but it was shown that the result can be written as the
convolution of the one-loop short distance coefficient with the one-loop
Brodsky-Lepage light-cone pion wavefunction~\cite{BL}. In the remainder of this
section we show that the allowed form of $Q_{\bf 0}^1$ reproduces this
convolution from one-loop graphs in the effective theory.

By parity $Q_{\bf 0}^{5}$ does not contribute, and we leave the discussion of
$Q_{\bf 8}^i$ to Ref.~\cite{proof}. In the matrix element for $B\to D\pi$
momentum conservation implies we can set $\bnP^\dagger\!-\! \bnP\!= 2E_\pi$,
similar to the example in section III.  Furthermore, with $m_\pi = 0$ the
kinematics of $B\to D\pi$ can be used to eliminate $v \mcdot v'$ and $E_\pi$ in
favor of $m_{c}$ and $m_{b}$. Therefore, we write $C^1_{\bf 0}(\mu, v\mcdot v',
{\bnP}, {\bnP^\dagger} ) \equiv C_{\bf 0}(\mu,{\bnP_+})$.  The example in this
section is nontrivial because $C_{\bf 0}$ still depends on $\bnP_+$.

The factors of $W$ contain an infinite set of terms with an arbitrary number of
collinear gluons emerging from the four-fermion operator.  The zero gluon
Feynman rule is
\begin{eqnarray}\label{feynrule1}
\begin{picture}(65,30)(1,1)
 \lower20pt  \hbox{\includegraphics[width=1in]{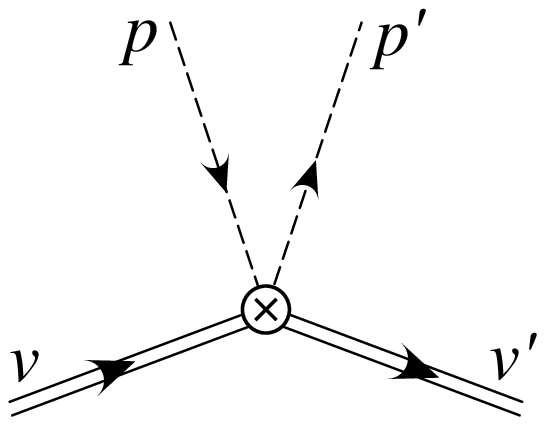}}
\end{picture}\qquad
  & =&\quad  i\,C_{\bf 0}\Big(\mu, \bar{n}\! \cdot\! (p + p') \Big)\ 
  \Gamma_h \otimes \Gamma_\ell \,. \\[-2pt] \nn
\end{eqnarray}
where $\Gamma_h$ and $\Gamma_\ell$ are the spin structures between the heavy and
light fields in $Q_{\bf 0}^1$.  At tree level, the matching onto
Eq.~(\ref{feynrule1}) is trivial, and $C_{\bf 0}^{(0)}(\mu,{\bnP_+}) =C_{\bf
0}^{\rm full}$. At one loop, matching at $\mu=m_b$ we find
\begin{eqnarray}\label{oneloop_C}
  C_{\bf 0}^{(1)} \Big(m_b,{\bnP_+}\Big) &=& C_{\bf 8}^{\rm full} 
  \frac{C_F\alpha_s}{2\pi N_c} \bigg\{ \left(3+2\ln\frac{u}{\bar{u}} 
  \right) \ln z^2 + f(u,z) + f(\bar u,1/z) -b \bigg\} + G(z) \,,
\end{eqnarray}
in agreement with Refs.~\cite{polwise,bbns2} (where $C_F=4/3$ and $N_c=3$).
Here $z = {m_c}/{m_b}$, $b$ is a scheme dependent constant, the function
$f(u,z)$ is given in Eq.~(85) of Ref.~\cite{bbns2}, and $G(z)$ contains the
standard hard contribution for the $B\to D$ form factor in
HQET~\cite{hardff}. In our formalism $u$ and $\bar u$ are operators, $u =
\frac{1}{2}[1 + {\bnP_+}/(2E_\pi)]$ and $\bar{u} = \frac{1}{2}[1 -
{\bnP_+}/(2E_\pi)]$.

The new information gained from the effective theory is that the Wilson
coefficients for terms with {\em any} number of collinear gluons are related to
one another. They are determined by expanding the exponential $W$ factors in the
operator $Q_{\bf 0}^1$ in Eq.~(\ref{fourfermi_effective}).  For example, for one
gluon the Feynman rule is
\begin{eqnarray}
\begin{picture}(65,30)(1,1) \label{feynrule2}
 \lower30pt  \hbox{\includegraphics[width=1in]{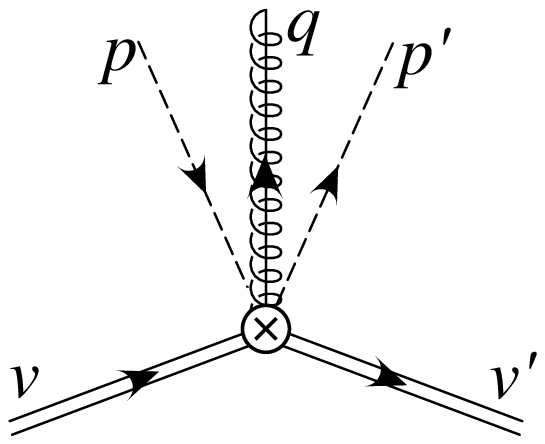}}
\end{picture}\ \
  & =& {ig\:\bn^\mu}\: \Gamma_h \otimes \Gamma_\ell\: T^A 
 \Bigg\{ \frac{C_{\bf 0}\Big(\mu,\bar{n}\!\cdot\!(p\!+\!p'\!+\! q)\Big) 
  - C_{\bf 0}\Big(\mu,\bar{n}\!\cdot\!(p\!+\!p'\!-\!q)\Big) }{\bn\mcdot q} 
  \Bigg\} \,, \\[5pt]\nn
\end{eqnarray}
with the same function $C_{\bf 0}$ as in Eq.~(\ref{feynrule1}). Note that with
tree level matching ($C_{\bf 0}=$ constant) the Feynman rules with $\ge 1$ gluon
vanish since $W^\dagger W=1$.

\begin{figure}[t]
\centerline{ \includegraphics[width=5in]{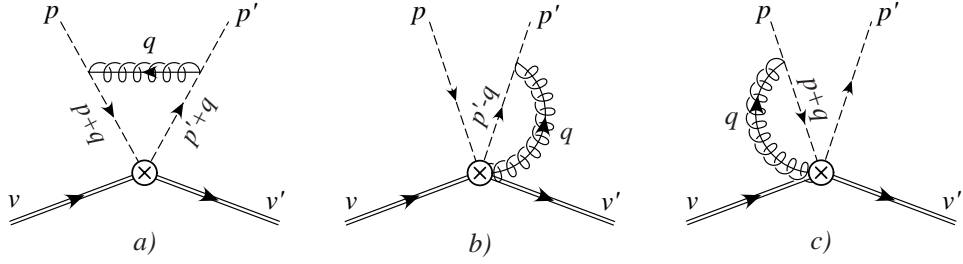} }
\caption{The collinear graphs contributing at one loop in the effective
theory.\label{bdpiloop}}
\end{figure}

Given the Feynman rules from Eqs.~(\ref{Lc}), (\ref{feynrule1}), and
(\ref{feynrule2}), we can obtain the one loop contribution to $B \to D \pi$ from
collinear gluon loops. The possible graphs are shown in
Fig.~\ref{bdpiloop}. Replacing $h_v^{(b)}$, $h_{v'}^{(c)}$, and
$\xi_{n,p'}^{(u)}\, \bar\xi_{n,p}^{(d)}$ with color singlet interpolating fields
with the correct spin structure, the spin and color traces can be performed to
give
\begin{eqnarray}
   i A_a &=& 2g^2 C_F \int \! \frac{d^d q}{(2\pi)^d}
      \: \frac{q_\perp^2}{(q+p)^2(q+p')^2 q^2} 
      \ C_{\bf 0}\Big(\mu,\bar{n}\! \cdot\! (p + p'+2q) \Big) \quad \mbox{+\, 
   \{$p_\perp$, $p^{\prime}_\perp$\} terms}   \,, \nonumber \\
   i A_b &=& -2 g^2 C_F \int \!\frac{d^d q}{(2\pi)^d}\: 
      \frac{\bn \!\cdot \!(p'-q)}{q^2(p'-q)^2}
      \ \frac{C_{\bf 0}\Big(\mu,\bar{n}\! \cdot\! (p + p') \Big) - 
       C_{\bf 0}\Big(\mu,\bar{n}\mcdot(p + p'-2q) \Big)}{\bn \! \cdot \!q}
      \,, \nonumber \\
   i A_c &=& -2 g^2 C_F \int \!\frac{d^d q}{(2\pi)^d}\: 
      \frac{\bn \!\cdot \!(p+q)}{q^2(p+q)^2}
      \ \frac{C_{\bf 0}\Big(\mu,\bar{n}\! \cdot\! (p + p' + 2q) \Big) - 
      C_{\bf 0}\Big(\mu,\bar{n}\mcdot(p + p') \Big)}{\bn \!\cdot\!q} \,,
\end{eqnarray}  
where a common prefactor is suppressed.  Adopting the notation of
Ref.~\cite{bbns2}, we let $q \to -q$ for the integration variable in $A_b$, and
let $p=-\bar u p_\pi$, $p'= u p_\pi$, $p_\pi= E_\pi n^\mu$, $\bn \! \cdot q
\equiv 2 \alpha E_\pi$, with $\bar{u} = 1-u$, and drop the $p_\perp$ and
$p_\perp'$ dependence.  The sum of the three graphs is then
\begin{eqnarray}\label{twoloopresult}
  i A_{abc} &=& g^2 C_F\! \int  \!\frac{d^dq}{(2\pi)^d}\Bigg\{\left[
   \frac{2(u\!+\!\alpha)}{\alpha}\frac{1}{(q+u p_\pi )^2 q^2} - 
   \frac{2(\bar{u}\!-\!\alpha)}{\alpha}\frac{1}{(q-\bar{u} p_\pi)^2 q^2} 
   \right] \Big[ T(u) \!-\! T(u\!+\!\alpha) \Big] \nonumber\\ 
  && \qquad\qquad\qquad\quad+ \frac{2 q_\perp^2}{(q+up_\pi)^2(q-\bar{u}p_\pi)^2 
   q^2}\: T(u\!+\!\alpha)\Bigg\}\,,
\end{eqnarray}
where 
\begin{eqnarray}
 T(x,\mu) \equiv C_{\bf 0}(\mu,(4x-2)E_\pi) \,.
\end{eqnarray}
Using the one loop result for the short distance coefficient, $T^{(1)}(x,\mu)$,
Eq.~(\ref{twoloopresult}) is exactly the result of the explicit two loop
calculation given in Eq.~(193) of Ref.~\cite{bbns2}. Invariance under the
collinear gauge transformation has enabled us to reproduce this result without
having to calculate two loop graphs.  

In Ref.~\cite{bbns2} it was shown that after performing the $q_\perp$ and
$n\mcdot q$ integrals in Eq.~(\ref{twoloopresult}) and adding the collinear wave
function contribution, $A_{abc}$ factors into the convolution of
$T^{(1)}(x,\mu)$ with the one-loop light cone pion wavefunction~\cite{BL},
$\Phi_{\pi}^{(1)}(x,\mu)$, i.e.
\begin{eqnarray}
  iA_{abc} \ + \ \ 
\begin{picture}(65,20)(1,1)
  \lower8pt \hbox{\includegraphics[width=1.0in]{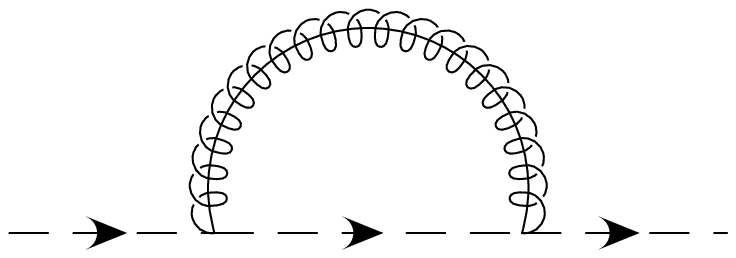}}
\end{picture} \ \ \
 &=&\  \int_{0}^{1} dx\: T^{(1)}(x,\mu)\: \Phi_{\pi}^{(1)}(x,\mu) \,.
\end{eqnarray}
However, our result in Eq.~(\ref{twoloopresult}) holds with $T(x)$ matched at
any order in perturbation theory, so the same steps reduce
Eq.~(\ref{twoloopresult}) to the convolution of the complete short distance
coefficient with $\Phi_{\pi}^{(1)}(x,\mu)$,
\begin{eqnarray}
  iA_{abc} \ + \ \ 
\begin{picture}(65,20)(1,1)
  \lower8pt \hbox{\includegraphics[width=1.0in]{WF}}
\end{picture} \ \ \
 &=&\  \int_{0}^{1} dx\: T(x,\mu)\: \Phi_{\pi}^{(1)}(x,\mu) \,.
\end{eqnarray}
This provides strong additional evidence for the validity of generalized
factorization~\cite{polwise,bbns} for $B\to D\pi$. It also illustrates the
utility of the effective theory description of collinear degrees of freedom.
Within this framework an all orders proof of generalized factorization is
presented in Ref.~\cite{proof}, where soft modes are included and the
convolution discussed above is extended to include the complete non-perturbative
pion light cone wavefunction.

\section{Conclusion} \label{sect_concl}

We have shown that gauge invariant operators in the collinear effective theory
can be constructed by satisfying four simple rules:
\begin{enumerate}

 \item[1)] Changes of variable on the labels of effective theory fields are
 allowed. \\[-20pt]

 \item[2)] In writing Feynman rules the label momenta must be conserved.\\[-20pt]
   
 \item[3)] Only operators invariant under collinear gauge transformations 
  (Eq.~(\ref{gt})) are allowed.\\[-20pt]

 \item[4)] Wilson coefficients are functions of the label operators $\bnP$ and
 $\bnP^\dagger$ and must be inserted in all possible locations in a field
 operator that are consistent with 3).\\[-15pt]
\end{enumerate} 
We have presented two simple examples by constructing the kinetic term for
collinear quarks and the heavy to light current. We then showed how the
formalism can be applied to the decay $B \to D \pi$. Only one color singlet
operator can be constructed which contributes to this decay, and we reproduced a
nontrivial two loop result in Ref.~\cite{bbns2}.  Furthermore, in the effective
theory the one-loop graphs with a single collinear gluon give a convolution of
the short distance coefficient at any order in perturbation theory with the
one-loop pion wavefunction.  Within this framework an all orders proof of
generalized factorization for $B\to D\pi$ is presented in Ref.~\cite{proof}.

\acknowledgements This work was supported in part by the Department of Energy
under the grant DOE-FG03-97ER40546 and by NSERC of Canada. We would like to
thank A.~Falk, B.~Grinstein, A.~Manohar, D.~Pirjol, I.~Rothstein, and M.~Wise
for comments on the manuscript.



\end{document}
\end